\documentclass[aps,prl,showpacs,twocolumn,amsmath,amssymb,superscriptaddress]{revtex4}
\usepackage{graphicx}
\usepackage{Jonasmacros}
\usepackage{bm}
\newcommand{\beqa}{\begin{eqnarray}}
\newcommand{\eeqa}{\end{eqnarray}}

\begin{document}
\preprint{}
\title{Vibrating Superconducting Island in a Josephson Junction}
\author{J. Fransson}
\email{Jonas.Fransson@fysik.uu.se}
\affiliation{Theoretical Division, Los Alamos National Laboratory,
Los Alamos, New Mexico 87545 }
\affiliation{Center for Nonlinear Studies, Los Alamos National Laboratory,
Los Alamos, New Mexico 87545 }
\affiliation{Department of Physics and Materials Science, Uppsala University, Box 530, SE-751 21\ Uppsala, Sweden}
\author{Jian-Xin Zhu}
\affiliation{Theoretical Division, Los Alamos National Laboratory,
Los Alamos, New Mexico 87545 }
\author{A. V. Balatsky}
\affiliation{Theoretical Division, Los Alamos National Laboratory,
Los Alamos, New Mexico 87545 }
\affiliation{Center for Integrated Nanotechnology, Los Alamos National Laboratory,
Los Alamos, New Mexico 87545 }
\begin{abstract}
We consider a combined nanomechanical-supercondcuting device that
allows the Cooper pair tunneling to interfere with the mechanical
motion of  the middle superconducting island. Coupling of mechanical
oscillations of a superconducting island between two superconducting
leads to the electronic tunneling generates a supercurrent
 that is modulated by the oscillatory motion of the island.
 This coupling produces alternating finite and vanishing
  supercurrent as function of the superconducting phases. Current peaks are
   sensitive to the superconducting phase shifts
  relative to each other. The proposed  device  may be used to study the nanoelectromechanical
   coupling in case of superconducting electronics.
\end{abstract}
\pacs{85.85.+j, 73.40.Gk, 82.25.Cp}
\date{\today}
\maketitle


Vibrational modes (vibrons) and spins possess dynamical degrees of freedom and they have a large impact on electron dynamics. Peaks and dips in the differential conductance of molecular electronics devices \cite{andres1996,reed1997,kergueris1999,hong2000,rosink2000,chen2000,porath2000,smit2002,reichert2002,aviram1998,langlais1999,park2000,park2002,zhitenev2002}
      may indicate strong effects from electron-vibron coupling. Steps in the differential
       conductance have been observed, both experimentally and theoretically,
       in STM based inelastic tunneling spectroscopy (IETS) around local vibrational
       mode on surfaces \cite{stipe1998,franssonIETS2007}. Effects from local vibrational
        modes on the conductance in molecular quantum dots and single electron transistor
         have also been investigated
         \cite{wingreen1989,lundin2002,zhu2003,mitra2004,flensberg2003,chtchelkatchev2004,kohler1993,engel2002,zhu2004,nussinov2005,zhu2006}.
         These studies may have implications on charge-based quantum information
         technology.

More recently the field has been moving towards
         incorporating superconducting electronics into
         nanoelectromechanical devices, e.g. see recent work
         \cite{schwab2006,meschke2006}. One example of superconducting electronics in combination with
         nanomechanical setup is a Cooper pair shuttle. The setup
          proposed originally \cite{gorelik2001} was aimed at a steady state description of the mechanical assisted Cooper pair tunneling. We extend the analysis of Cooper pair shuttle
          to consider
          possible resonances between mechanical motion and ac
          Josephson effect. We here address
         dynamical aspects of this device where we focus on time domain in the presence of ac signatures
         in the electric current.

In this Letter we study Josephson tunneling in a system with a
mechanically
moving superconducting island between two
 superconducting leads. Mechanical
 oscillations of the island couples to the
 electronic tunneling. This coupling gives
 rise to an oscillator modulation of the
  Josephson current, such that the Fourier
   spectrum in general exhibits current
   peaks at $\omega_J\pm\tilde\omega_0$, $\omega_J\pm2\tilde\omega_0$,
   and $2\omega_J$, where $\omega_J=$eV and $\tilde\omega_0$
    are the Josephson and renormalized oscillator frequencies,
     respectively. In addition, the zero bias
      supercurrent peaks at the
       frequencies $0$, $\pm\tilde\omega_0$,
        and $\pm2\tilde\omega_0$, however, the
         peaks at $0$ and $\pm2\tilde\omega_0$,
          or $\pm\tilde\omega_0$ vanish for
          certain combinations of the
           superconducting phases in
           the three components of the
            system. 

\begin{figure}[b]
\begin{center}
\includegraphics[width=8.5cm]{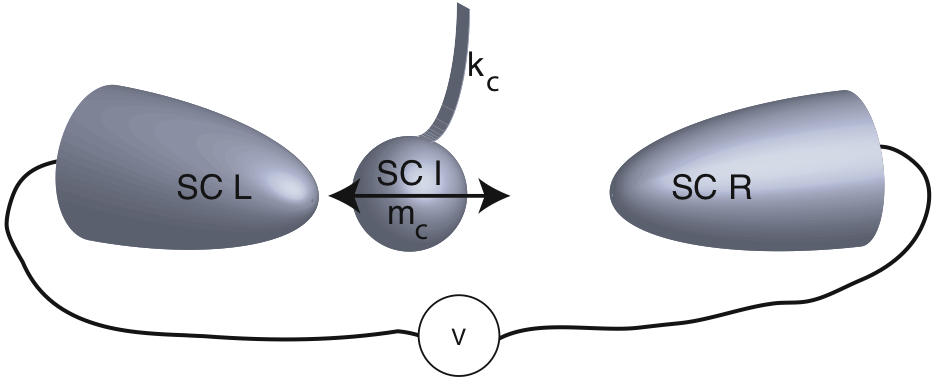}
\end{center}
\caption{Schematic view of the mechanically and
electronically coupled superconducting island
(SC I) to the superconducting leads (SC L and SC R).
The cantilever superconducting island is modeled as
a harmonic oscillator with spring constant $k_c$ and
mass $m_c$, placed between the infinitely massive
 superconducting leads. The device is biased with
 voltage $V$.}
\label{fig-system}
\end{figure}

The physical system we are
consider consists of a
superconducting island
between two superconducting
leads, where the island is
exposed to mechanical vibrations
 modeled with Hook's law force
 constant $k_c$, see Fig. \ref{fig-system}. A bias voltage is applied
 across the junction. The Hamiltonian for
  the system is given by
\begin{equation}
\Hamil=\Hamil_L+\Hamil_I+\Hamil_R+\Hamil_T.
\label{eq-model}
\end{equation}
Here the leads $(L,R)$ and the island $(I)$ are
described by the BCS Hamiltonians as $\Hamil_{L(I,R)} =\sum_{p(k,q)\sigma}\dote{p(k,q)}\cdagger{p(k,q)}\cc{p(k,q)}+\sum_{p(k,q)}[\Delta_{L(I,R)}\csdagger{p(k,q)\up}\csdagger{-p(-k,-q)\down}+H.c.]$.
We denote the creation (annihilation) operators by $\cdagger{p(k,q)}\
(\cc{p(k,q)})$, where the subscript $p\, (k,q)$ is momenta in the  leads and island,
whereas $\sigma=\up,\down$ is a spin index. Finally, $\dote{p(k,q)}$ and
$\Delta_{L(I,R)}$ are the single-electron energies and pair potential (gap
function), respectively. 
Without loss of generality, we assume that the
superconductors are of a conventional spin-singlet $s$-wave pairing symmetry, and
consider the Josephson tunneling at zero temperature. The last
assumption implies that the temperature is low compared to relevant
superconducting gaps, e.g. $T\ll\Delta_{L(I,R)}$.  To avoid
thermal damping of an oscillator we also require low enough $T\sim$
10 | 100 mK. The last term in Eq. (\ref{eq-model}) describes the
tunneling between the leads and the island, i.e.
$\Hamil_T=\sum_{pk\sigma}[T_{pk}\cdagger{p}\cc{k}+H.c.]+\sum_{qk\sigma}[T_{qk}\cdagger{q}\cc{k}+H.c.]$, where the tunneling matrix elements $T_{p(q)k}$
transfers electrons through insulating barriers
between the leads and the island. The local
vibrational mode of the island is in the
linear coupling regime given by
\begin{equation}
T_{pk}=T_{pk}^{(0)}(1+\alpha_L u),\ \quad T_{qk}=T_{qk}^{(0)}(1+\alpha_R u),
\end{equation}
where $\alpha_{L(R)}$ describes the coupling
between the tunneling electrons and the
 vibrational mode corresponding to the
  left (right) tunnel junction. The
  quantity $u$ is the displacement operator
  for the oscillator. The tunneling matrix element $T_{pk}$ is exponential in displacement $u$, thus, the assumed linear coupling is a good approximation for small $u$. This allows evaluation of $\alpha_{L,R}$ in terms of the tunneling matrix elements and their distance dependence. 
  We assume here a very general equilibrium
   geometry with  no particular symmetry
  being required. The
  equilibrium point $(u=0)$ of the mechanical
   oscillator placed within the junction  could be placed anywhere in between the leads. The
    energy associated with the vibrational mode, $\omega_0=\sqrt{k_c/m_c}\sim10^{-1} - 10^{-6}\ \text{eV}$, is much smaller than the typical electronic energy on the order of 1 eV, the mechanical oscillations are very slow compared to the time scale of the electronic processes. This allows us to apply the Born-Oppenhiemer approximation to treat the electronic degrees of freedom as if the local oscillator is static at every instantaneous location.

The current is derived using standard methods. For a given bias voltage $eV=\mu_L-\mu_R$, where $\mu_\chi$, $\chi=L,R$ is the chemical potential of the left ($L$) and right ($R$) lead, the Josephson current between the lead $\chi$ and the island is given by \cite{zhu2006} (setting $\hbar=1$)
\begin{eqnarray}
I_S^\chi(t)&=&J_S^\chi(\omega_\chi)[1+\alpha_\chi u]^2
   \sin{(\omega_J^\chi t+\phi_\chi)}
\nonumber\\&&
   -\Gamma_S^\chi(\omega_\chi)
   [1+\alpha_\chi u]\alpha_\chi\dot{u}
   \cos{(\omega_J^\chi t+\phi_\chi)},
\label{eq-jcurrent}
\end{eqnarray}
in applying the local approximation $u(t')\simeq u(t)+(t'-t)\dot{u}(t)$. Here we have introduced the phase difference $\phi_\chi$ between the lead $\chi$ and the island, and the Josephson frequency $\omega_J^\chi=2(\mu_\chi-\mu_I)$. In Eq. (\ref{eq-jcurrent}), $J_S^\chi$ is the amplitude of the Josephson current when the cantilever is frozen ($u=0$), given by
\begin{eqnarray}
J_S^\chi(\omega_\chi)&=&e\sum_{n\in\chi, k}
   \frac{|T^{(0)}_{n k}|^2|\Delta_\chi\Delta_I|}{E_n E_k}
   \biggl(\frac{1}{\omega_\chi+E_n+E_k}
\nonumber\\&&
   -\frac{1}{\omega_\chi-E_n-E_k}\biggr),
\end{eqnarray}
where $\Delta_\chi$ and $\Delta_I$ is the superconducting gap in the lead $\chi$ and island, respectively, whereas the quasi-particle energies $E_n=\sqrt{(\dote{n}-\mu_\chi)^2+|\Delta_\chi|^2}$ and $E_k=\sqrt{(\dote{k}-\mu_I)^2+|\Delta_I|^2}$. The second contribution to the Josephson current in Eq. (\ref{eq-jcurrent}) has the amplitude
\begin{eqnarray}
\Gamma_S^\chi(\omega_\chi)&=&e\sum_{n k}
   \frac{|T_{n k}^{(0)}|^2|\Delta_\chi\Delta_I|}{E_n E_k}
   \biggl(\frac{1}{(\omega_\chi+E_n+E_k)^2}
\nonumber\\&&
   -\frac{1}{(\omega_\chi-E_n-E_k)^2}\biggr).
\end{eqnarray}
Using that $I_S^R=-I_S^L$ for stationary bias voltages we write the total Josephson current $I_S=I_S^L=(I_S^L-I_S^R)/2$.

The Hamiltonian $\Hamil_J$ for the Josephson energy \cite{zhu2006} is derived by requiring that the derivative of $\Hamil_J$ with respect to $\phi_\chi$ yields the supercurrent given in Eq. (\ref{eq-jcurrent}). We find that
\begin{eqnarray}
\Hamil_J&=&\frac{1}{2}\sum_{\chi=L,R}\biggl(E_J^\chi(1+\alpha_\chi u)^2
   [1-\cos(\omega_J^\chi t+\phi_\chi)]
\nonumber\\&&
   -\frac{1}{2e}\Gamma_S^\chi(1+\alpha_\chi u)\alpha_\chi\dot{u}
       \sin(\omega_J^\chi t+\phi_\chi)\biggr)
\label{eq-Heff}
\end{eqnarray}
where $E_J^\chi=J_S^\chi/(2e)$.
 This
     effective Hamiltonian   captures
      the back action effect and
       reflects how the electronic
       degrees of freedom influence
       the mechanical oscillator.
       The classical motion of the
       shuttling island is  described by the Hamiltonian $\Hamil_{osc}=p^2/(2m)+k_cu^2/2+\Hamil_J$, which results in the classical equation of motion
\begin{equation}
m_c\ddot{u}+[\gamma_S(t)+\gamma_N]\dot{u}+k_cu=F(t).
\end{equation}
Here, the driving force
\begin{eqnarray}
F(t)&=&-\sum_{\chi=L,R} E_J^\chi\alpha_\chi(1+\alpha_\chi u)\biggl[1
   -\Bigl(1+\frac{\omega_J^\chi\Gamma_S^\chi}{4eE_J^\chi}\Bigr)
\nonumber\\&&\times
       \cos(\omega_J^\chi t+\phi_\chi)\biggr],
\end{eqnarray}
whereas the time-dependent damping factor $\gamma_S(t)+\gamma_N$ contains the superconducting contribution 
\begin{equation}
\gamma_S(t)=-\frac{1}{2e}\sum_{\chi=L,R}
   \Gamma_S^\chi\alpha_\chi^2\sin(\omega_J^\chi t+\phi_\chi).
\end{equation}
due to damping energy in and out of the superconducting carriers, and the external damping $\gamma_N$. We assume that the damping into non-superconducting degrees of freedom is suppressed by the factor $\exp{(-\Delta_{L,R}/T)}$, as one would need to damp energy into normal excitations. 

In order to make next steps analytically, we assume the system to be in the weak coupling limit. Moreover, since $\Gamma_S^\chi\ll J_S^\chi$ the main physics is captured by neglecting the terms in the damping and driving force proportional to $\alpha_\chi\Gamma_S^\chi$ and $\alpha_\chi^2$. The motion of the island is then given by
\begin{eqnarray}
u(t)&=&u_0\sin{(\tilde\omega_0t+\delta_0)}e^{-\gamma_Nt/2m_c}
	-\sum_\chi\frac{\alpha_\chi E^J_\chi}{k_c}
	[1
	-{\cal A}_\chi
\nonumber\\&&\times
	\{\cos{(\omega_J^\chi t+\phi_\chi)}
	+{\cal B}_\chi\sin{(\omega_J^\chi t+\phi_\chi)}
	\}],
\end{eqnarray}
where
\begin{eqnarray}
{\cal A}_\chi&=&	
				\frac{1-(\omega_J^\chi/\omega_0)^2}
						{[1-(\omega_J^\chi/\omega_0)^2]^2+(\gamma_N\omega_J^\chi/k_c)^2}
\\
{\cal B}_\chi&=&\frac{\gamma_N\omega_J^\chi}{k_c}\frac{1}{1-(\omega_J^\chi/\omega_0)^2}.
\end{eqnarray}
The first term describes the unperturbed motion of the island around its equilibrium position, where $u_0$ and $\delta_0$ are to be determined by the initial conditions, whereas $\dote{\pm}=-\gamma_N/2m\pm i\tilde\omega_0$ with eigenfrequencies $\tilde\omega_0=\sqrt{\omega_0^2-[\gamma_n/2m]^2}$ of the mechanical oscillations. The other terms arise due to the coupling between the mechanical and electronic degrees of freedom. Using this expression for the mechanical motion of the island, we obtain the Josephson current approximately as
\begin{eqnarray}
\vphantom{\sum_k}
\lefteqn{
I_S^\chi(t)=J_S^\chi\bigl[1
   +2\tilde\alpha_\chi\sin(\tilde\omega_0t+\delta_0)e^{-\gamma_Nt/2m_c}
}
\nonumber\\&&
   +\tilde\alpha_\chi^2\sin^2(\tilde\omega_0t+\delta_0)e^{-\gamma_Nt/m_c}
   -2\tilde\alpha_\chi
       \sum_{\chi'=L,R}\frac{\tilde\alpha_{\chi'}}{K_{\chi'}}(1
\nonumber\\&&
\vphantom{\sum_k}
	-{\cal A}_{\chi'}
	\{\cos(\omega_J^{\chi'}t+\phi_{\chi'})
		+{\cal B}_{\chi'}
		\sin(\omega_J^{\chi'}t+\phi_{\chi'})
		\})\bigr]
\nonumber\\&&\times
\vphantom{\sum_k}
	\sin(\omega_J^\chi t+\phi_\chi),
\label{eq-cvJ}
\end{eqnarray}
where $\tilde\alpha_\chi=u_0\alpha_\chi$ and $K_\chi=k_cu_0^2/E_\chi$.

We now consider a few limiting cases in the symmetrically biased system, such that $\omega_L=-\omega_R=\omega_J\ (=eV)$, and we assume that $T_{pn}^{(0)}=T_{qn}^{(0)}$ and $|\Delta_L|=|\Delta_R|$. Then, $J_S^L=J_S^R=J_S$, hence, we set $E_J^\chi=E_J$ and $K_\chi=K$. Moreover, ${\cal A}_L={\cal A}_R={\cal A}$, and ${\cal B}_L=-{\cal B}_R={\cal B}$, and we also assume that $\alpha_R=-\alpha_L=-\alpha$. Using Eq. (\ref{eq-cvJ}) we find that the total current under those circumstances can be written as
\begin{widetext}
\begin{eqnarray}
\frac{2I_S(t)}{J_S}&=&
	[1+\tilde\alpha^2\sin^2(\tilde\omega_0t+\delta_0)e^{-\gamma_Nt/m_c}]
	    [\sin(\omega_Jt+\phi_L)
	    	+\sin(\omega_Jt-\phi_R)]
	+2\tilde\alpha\sin(\tilde\omega_0t+\delta_0)
	    [\sin(\omega_Jt+\phi_L)
\nonumber\\&&
\vphantom{\frac{1}{2}}
	    	-\sin(\omega_Jt-\phi_R)]e^{-\gamma_Nt/2m_c}
	+\tilde\alpha^2{\cal A}
			[\sin2(\omega_Jt+\phi_L)
				+\sin2(\omega_Jt-\phi_R)
				-2\sin(2\omega_Jt+\phi_L-\phi_R)
\nonumber\\&&
\vphantom{\frac{1}{2}}
	+{\cal B}
		\{2-\cos2(\omega_Jt+\phi_L)-\cos2(\omega_Jt-\phi_R)
			+2\cos(2\omega_Jt+\phi_L-\phi_R)
				-2\cos(\phi_L+\phi_R)\}
				]/K
\label{eq-cvJappr}
\end{eqnarray}
\end{widetext}
In the case where the phases $\phi_\chi=0$ the Josephson current thus reduces to
\begin{equation}
\frac{2I_S(t)}{J_S}=2\biggl(1
	+\tilde\alpha^2\sin^2(\tilde\omega_0t+\delta_0)e^{-\gamma_Nt/m_c}
	\biggr)
   \sin\omega_Jt.
\end{equation}
\begin{figure}[t]
\begin{center}
\includegraphics[width=8.5cm]{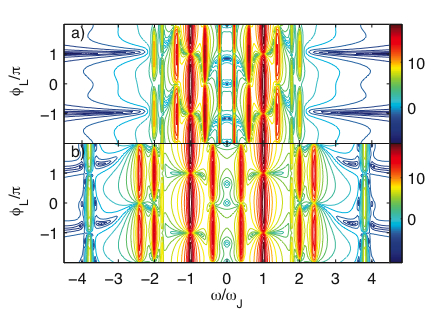}
\end{center}
\caption{(Color online) Fourier transform of the Josephson current $\log_{10}|2I_S(\omega)/J_S|^2$ in the symmetrically biased and undamped system as a function of the frequency $\omega$ and the phase $\phi_L$, for a fixed phase $\phi_R=0$. Here we have taken $\tilde\alpha=0.1$, $K=0.6$, and the oscillating frequencies a) $\omega_0=0.4\omega_J$ and b) $\omega_0=1.4\omega_J$.}
\label{fig-symIS}
\end{figure}
In the undamped, or very weakly damped case, this expression predicts that there will be a dc component to the Josephson current whenever the bias voltage matches the Shapiro step value $2\tilde\omega_0$. On the other hand, there is no dc component to the current at the value $\tilde\omega_0$ as is predicted for a single tunnel junction in Ref. ~\cite{zhu2006}. The motion of the island between the two leads, clearly, has a compensating effect in the sense that the coupling of the mechanical oscillator and the tunneling electrons cancel their respective side peaks at $\omega_J\pm\tilde\omega_0$.  These observations open up possibilities to excite cantilever motion by generating the Josephson current with appropriate frequencies. The features predicted from Eq. (\ref{eq-cvJappr}) at $\phi_\chi=0$, that is, the absence of Josephson current at $\omega_J\pm\tilde\omega_0$ and $2\tilde\omega_J$, while there is a finite Josephson current at $\omega_J\pm2\tilde\omega_0$, are clearly illustrated in Fig. \ref{fig-symIS}. These plots show the Fourier transform of the Josephson current $\log_{10}|2I_S(\omega)/J_S|^2$, in the undamped case $(\gamma_N=0$), as function of the phase $\phi_L$ and frequency $\omega$, for the fixed phase $\phi_R=0$ and two different frequencies $\omega_0$.

\begin{figure}[b]
\begin{center}
\includegraphics[width=8.5cm]{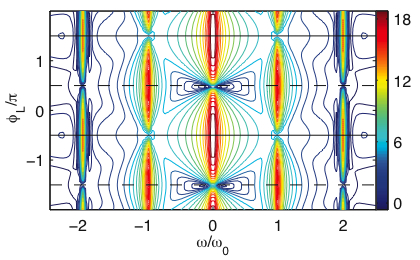}
\end{center}
\caption{(Color online) Zero bias Josephson current $\log_{10}|2I_S(\omega)/J_S|^2$ in the symmetrically biased and undamped system as a function of the frequency $\omega$ and the phase $\phi_L$, for a fixed phase $\phi_R=\pi/2$. Here we have taken $\omega_0=1$, while other parameters are as Fig. \ref{fig-symIS}.}
\label{fig-symISzb}
\end{figure}

More interesting is the zero bias Josephson current found from Eq. (\ref{eq-cvJappr}). Assuming that the phases $\phi_L=\phi_R=\phi$ we obtain
\begin{eqnarray}
\frac{2I_S(t)}{J_S}&=&4\tilde\alpha
	\sin(\tilde\omega_0t+\delta_0)\sin\phi \, e^{-\gamma_Nt/2m_c}
\end{eqnarray}
Hence, there is an ac component which is modulated by the eigenfrequency $\tilde\omega_0$ of the moving island. The dashed lines in Fig. \ref{fig-symISzb} signify equal phases $\phi_\chi=\phi$, at which only the ac component which is modulated by $\tilde\omega_0$ is finite.

In the case of opposite phases, $\phi_L=-\phi_R=\phi$, we find the Josephson current
\begin{equation}
\frac{2I_S(t)}{J_S}=2\biggl(1+\tilde\alpha^2\sin^2(\tilde\omega_0t+\delta_0)
	e^{-\gamma_Nt/m_c}\biggr)\sin\phi
\end{equation}
In contrast to the previous case, in this case there is a dc component depending on the phase $\phi$. Moreover, opposite phases give rise to an ac component which is modulated by twice the eigenfrequency ($2\tilde\omega_0$) of the moving island. Opposite phases are signified by solid lines in Fig. \ref{fig-symISzb}, showing a finite dc component to the Josephson current, along with finite ac components at $\pm2\tilde\omega_0$.

For an experiment of this type, taking $|\Delta|\sim10$ meV as relevant to MgB$_2$ \cite{stipe2001} provides a sufficiently small mechanical damping. By acquiring a vibrational frequency of $\omega_0/2\pi\sim1$ GHz \cite{huang2003}, it should be possible to tune the Josephson frequency such that $\omega_0/\omega_J\gtrsim0.1$. A coupling strength $\alpha/\omega_0\sim10^{-1} - 10^{-3}$ \cite{irish2005} would be sufficient to enable read-out of our predictions.


This work has been supported by US DOE, LDRD and BES,
and was carried out under the auspices of the NNSA of
 the US DOE at LANL under Contract No. DE-AC52-06NA25396.

\end{document}